\documentstyle[aps,prl,twocolumn,epsfig]{revtex}
\begin{document} 

\title{ Massive Quark Propagator in the Color-superconducting Phase}
\author{ HUANG Mei $^{1}$, ZHUANG Peng-Fei $^{1}$, CHAO Wei-Qin $^{2}$ \\
$^1$ Physics Department, Tsinghua University, Beijing 100084, China \\
huangmei@mail.tsinghua.edu.cn  ~ Tel: 62772694 \\
zhuangpf@mail.tsinghua.edu.cn ~ Tel: 62772696 \\[2mm]
$^2$ CCAST, Beijing 100080, China \\ 
Institute of High Energy Physics, CAS, Beijing 100039, China \\
chaowq@hp.ccast.ac.cn ~ Tel: 62569357}

\maketitle

\begin{abstract}
\small
A more general expression for the quark propagator including both
chiral and diquark condensates has been derived by using energy projectors. 
This makes it possible to study the phase transition from hadron phase  
to color superconductivity phase in the moderate baryon density region
by using Feynman diagrammatic method or Green-function method. 
\end{abstract}
{\bf PACS: 12.38.Aw; 11.10.Wx; 21.65.+f} \\

While the standard BCS (Bardeen Cooper and Schrieffer)
theory predicted the existence of color-superconducting 
phase at high baryon density twenty years ago \cite{in77}\cite{in78}, 
QCD (Quantum Chromodynamics) phase transitions along the baryon density direction 
attracts much attention after the appearance of \cite{raja1} and \cite{schafer}.
The authors of the two papers found that due to the non-perturbative effects, 
the color-superconducting gap can be of the order of $100 {\rm MeV}$,  which 
is two orders larger than early perturbative estimates \cite{love}. 

The color superconductivity with two massless 
flavors and the color-flavor-locking (CFL) phase with three degenerate massless 
quarks \cite{npb537} have been widely discussed from first principle QCD 
calculations. For review of the rich QCD phase structure at high baryon 
density, see \cite{raja2} and references therein. Usually, when the Green-function or 
Feynman diagrammatic method is used at the asymptotic densities  
\cite{rischke1} \cite{rischke2},  current quark mass or chiral condensate is 
not necessary to be considered because it can be neglected comparing with
the very high Fermi surface.
At less-than-asymptotic densities, the importance of non-zero quark mass 
corrections to the pure CFL phase and to the meson excitation was discussed in 
\cite{massraja1}, \cite{massschafer1}, \cite{massraja2} - \cite{rho2}, 
and the structure of the mass term was recently 
investigated in effective theories in \cite{massschafer4}. 
The anti-particles were decoupled from the system, and the quark mass corrections 
were treated perturbatively by expanding the quark mass order by order \cite{rho2}
\cite{massschafer4}. 

However, for physical applications, we are more interested in moderate baryon density 
region which may be related to the neutron 
stars and even in very optimistic cases --  to heavy-ion collisions.
In this region, the usual way is to work out the gap equations 
from the thermodynamic potential 
\cite{raja3}-\cite{ebert3} by using the variational methods.
To work out the phase structure from hadron gas to color superconductivity, 
one should deal with the chiral condensate and diquark condensate 
simultaneously. Since the chiral condensate contributes a dynamic quark mass, 
it becomes unreasonable to treat the quark mass term perturbatively 
like in \cite{rho2}. 

In this paper, we are trying to evaluate the quark propagators 
including chiral condensate and diquark condensate simultaneously.

At moderate baryon density, the four-fermion interaction models 
are usually used. With only scalar, pseudoscalar mesons and scalar diquark involved, 
the Lagrangian density has the form
\begin{eqnarray}
\label{lagr}
{\cal L} & = & {\bar q}(i\gamma^{\mu}\partial_{\mu}-m_0)q + 
   G_S[({\bar q}q)^2 + ({\bar q}i\gamma_5{\bf {\vec \tau}}q)^2 ] \nonumber \\
 & & +G_D[(i {\bar q}^C  \varepsilon  \epsilon^{\rho} \gamma_5 q )
   (i {\bar q} \varepsilon \epsilon^{\rho} \gamma_5 q^C)],
\end{eqnarray}
where $q^C=C {\bar q}^T$, ${\bar q}^C=q^T C$ 
are charge-conjugated spinors, $C=i \gamma^2 \gamma^0$ is the charge 
conjugation matrix (the superscript $T$ denotes the transposition operation),
$m_0$ is the current quark mass, the quark field $q \equiv q_{i\alpha}$ is a flavor 
doublet and color triplet as well as a four-component Dirac spinor with $i=1,2$, 
$\alpha=1,2,3$. ${\vec \tau}=(\tau^1,\tau^2,\tau^3)$ are Pauli matrices in the flavor 
space, and $(\varepsilon)^{ik} \equiv \varepsilon^{ik}$,
$(\epsilon^b)^{\alpha \beta} \equiv \epsilon^{\alpha \beta b}$ are totally 
antisymmetric tensors in flavor and color spaces.  
  
After bosonization by introducing the bosonic fields $ \Delta^{\pm}, \sigma, \pi$,
and the 8-component spinors 
$\Psi$ and $ \Psi_3 $ for the quarks participating and not participating in
the diquark condensate, 
\begin{equation}
\Psi =\left(\begin{array}{c} 
                    Q \\
                    Q^C 
                   \end{array}
            \right), ~~\Psi_3 =\left(\begin{array}{c} 
                    q_3 \\
                    q_3^C
                   \end{array}
            \right), ~~ 
\end{equation}
the partition function can be evaluated as
\begin{eqnarray}
\label{part}
{\cal Z} & = & N'{\rm exp} \{- \int_0^{\beta} d \tau \int d^3{\vec x} ~ 
[\frac{\sigma^2}{4 G_S}
      +\frac{\Delta^{+}\Delta^{-}}{4 G_D}] \} \nonumber \\
   & \times &  \int[d \Psi_3]{\rm exp}\{\frac{1}{2}\sum_{n,{\vec p}}  
 ~ {\bar \Psi}_3\frac{G_0^ {-1}}{T}\Psi_3 \}  \nonumber \\
  & \times & \int[d \Psi]{\rm exp} \{\frac{1}{2}  \sum_{n,{\vec p}} ~
 {\bar \Psi}\frac{{\rm G}^{-1}}{T}  \Psi \},
\end{eqnarray}
where we have assumed there was no pion condensate and  
\begin{equation}
{\rm G_0}^{-1} = 
    \left( 
          \begin{array}{cc}
            \left[ G_0^{+} \right]^{-1}  &  0 \\  
            0 &  \left[ G_0^{-} \right]^{-1} 
              \end{array}  
             \right)
\end{equation}
with 
\begin{eqnarray}
[G_0^{\pm}]^{-1} & = &
(p_0 \pm \mu) \gamma_0 -{\vec \gamma}\cdot {\vec p} -m, \nonumber \\
m & = & m_0+  \sigma , \ \ p_0=i \omega_n,
\end{eqnarray}
is the quark propagator in the normal phase, and
\begin{equation}
{\rm G}^{-1} = 
    \left( 
          \begin{array}{cc}
            \left[ G_0^{+} \right]^{-1}  &  \Delta^{-} \\  
            \Delta^{+}  &  \left[ G_0^{-} \right]^{-1} 
              \end{array}  
             \right) 
\end{equation}
with 
\begin{eqnarray}
\label{delta}
\Delta^{-} & = & -i \Delta \varepsilon  \epsilon^{b} \gamma_5, \ \ 
\Delta^{+} = -i \Delta^{*} \varepsilon  \epsilon^{b} \gamma_5, 
\end{eqnarray}
is the quark propagator in the color-breaking phase. Here
we have neglected the color, flavor indices.

By solving $1={\rm G}^{-1}\, {\rm G}$, the Nambu-Gorkov propagator can be expressed as 
\cite{rischke1}\cite{rischke2}
\begin{equation} 
\label{S}
{\rm G} = \left( \begin{array}{cc}
  G^{+} &   \Xi^{-}  \\
 \Xi^{+}  & G^{-}
\end{array}  \right) \,\, 
\end{equation}
with the off-diagonal elements
\begin{eqnarray}
\label{cosi}
\Xi^{\pm} \equiv -  G^{\mp} \, \Delta^{\pm} \, G_0^{\pm}
\end{eqnarray}
and the diagnal element  
\begin{equation}
\label{gpm}
G^{\pm} \equiv \left(\left( G_0^{\pm} \right)^{-1} - \Sigma^{\pm} 
\right)^{-1} \,\,\, , \,\,\,\, \Sigma^{\pm} \equiv \Delta^{\mp}
\, G_0^{\mp}\, \Delta^{\pm} \,\, .
\end{equation}
Here all the elements are 4-momentum $p^\mu$ dependent.

In the very high baryon density region, the quark mass $m$ can be neglected.  
By using the energy projectors for massless fermions 
\begin{eqnarray}
\Lambda_{\pm}({\vec p})=\frac{1}{2}(1\pm\frac{\gamma_0 {\vec \gamma}\cdot{\vec p}}{|{\vec p}|}),
\end{eqnarray}
the propagator had been evaluated in \cite{rischke1} and \cite{rischke2} as  
\begin{eqnarray}
\label{limitm}
G^{\pm}   &=&  \frac{p_0-\epsilon_p^{\pm}}{p_0^2-\left({\epsilon_p^{\pm}}\right)^2
-\Delta^2}\gamma_0\Lambda_{+}
+ \frac{p_0+\epsilon_p^{\mp}}{p_0^2-\left({\epsilon_p^{\mp}}\right)^2
-\Delta^2}\gamma_0\Lambda_{-}\ ,\nonumber\\
\Xi^{\pm} &=& \frac{\mp\Delta \gamma_5}{p_0^2
-\left({\epsilon_p^{\pm}}\right)^2-\Delta^2}\Lambda_{+}
+\frac{\mp\Delta\gamma_5}{p_0^2-\left({\epsilon_p^{\mp}}\right)^2
-\Delta^2}\Lambda_{-}\ ,
\end{eqnarray}
with $\epsilon_p=|{\vec p}|$ and $\epsilon_p^{\pm}= |{\vec p}| \pm \mu$.

In our case at moderate baryon density, the quark mass is constituent mass, which includes 
the contribution from the chiral condensate and can not be neglectd.
We need to use the energy projectors for massive free particles,
\begin{eqnarray}
\label{lpm}
\Lambda_{\pm}({\vec p})=\frac{1}{2}(1\pm\frac{\gamma_0({\vec \gamma}\cdot{\vec p}+m)}{E_p}),
\end{eqnarray}
with the quark energy $E_p=\sqrt{{\vec p}^2+m^2}$. Besides the basic properties
\begin{eqnarray}
\label{pro}
& & \Lambda_{\pm}({\vec p})\Lambda_{\pm}({\vec p})=\Lambda_{\pm}({\vec p}), \nonumber \\
& & \Lambda_{\pm}({\vec p})\Lambda_{\mp}({\vec p})=0, \nonumber \\
& & \Lambda_{+}({\vec p})+\Lambda_{-}({\vec p})=1\ , 
\end{eqnarray}
we can derive the following transformation properties 
\begin{eqnarray}
\label{gamma}
&  & \gamma_0 \Lambda_{\pm}({\vec p}) \gamma_0=\tilde \Lambda_{\mp}({\vec p}), \nonumber \\
&  & \gamma_5 \Lambda_{\pm}({\vec p}) \gamma_5=\tilde \Lambda_{\pm}({\vec p}),
\end{eqnarray}
where we have defined  
\begin{eqnarray}
\label{tilde}
\tilde \Lambda_{\pm}({\vec p})=
\frac{1}{2}(1\pm\frac{\gamma_0({\vec \gamma}\cdot{\vec p}-m)}{E_p})\ .
\end{eqnarray}
$\tilde \Lambda_\pm$ have different sign in mass term comparing with $\Lambda_{\pm}$.  
From Eqs. (\ref{lpm}) and (\ref{tilde}), the $\gamma_0$ transformation changes the signs of 
energy and mass in $\Lambda_\pm$, but the $\gamma_5$ transformation changes 
only the sign of mass. 
It can be proven that $\tilde\Lambda_{\pm}({\vec p})$ are energy projectors too.
In the limit of quark mass $ m \rightarrow 0$, we have 
$E_p=|{\vec p}|$, and
\begin{eqnarray} 
\tilde \Lambda_{\pm}({\vec p})=\Lambda_{\pm}({\vec p})
=\frac{1}{2}(1\pm\frac{\gamma_0{\vec \gamma}\cdot{\vec p}}{|\vec p|})\ .
\end{eqnarray}

Using the energy projectors we can re-express the normal quark propagator elements as 
\begin{eqnarray}
\label{mass0}
G_0^{\pm}& = & 
\frac{1}{\gamma_0(p_0-E_p^{\mp})\Lambda_{+}+\gamma_0(p_0+E_p^{\pm})\Lambda_{-}} \nonumber \\
& = & \frac{1} {(\gamma_0(p_0-E_p^{\mp})\Lambda_{+}+\gamma_0(p_0+E_p^{\pm})\Lambda_{-})} \times
\nonumber \\
& & \frac{\gamma_0(p_0-E_p^{\mp})\tilde \Lambda_{+}+\gamma_0(p_0+E_p^{\pm})\tilde\Lambda_{-}}
{(\gamma_0(p_0-E_p^{\mp})\tilde \Lambda_{+}+\gamma_0(p_0+E_p^{\pm})\tilde \Lambda_{-})} 
\nonumber \\
& = &  \frac{\gamma_0\tilde \Lambda_{+}}{p_0+E_p^{\pm}} + 
\frac{\gamma_0\tilde\Lambda_{-}}{p_0-E_p^{\mp}}
\end{eqnarray}
with $E_p^\pm = E_p \pm \mu$.

Taking the diquark condensate into account, we derive the elements of the Nambu-Gorkov propagator
\begin{eqnarray}
\label{gpmagain}
G^{\pm} & = & \left(\left(G_0^{\pm}\right)^{-1} - \Delta^{\mp}
\, G_0^{\mp}\, \Delta^{\pm} \right)^{-1} \nonumber \\
& = & [ \gamma_0(p_0-E_p^{\mp})\Lambda_{+} + \gamma_0(p_0+E_p^{\pm})\Lambda_{-} \nonumber \\
& & + \Delta \gamma_5 (\frac{\gamma_0\tilde\Lambda_{+}}{p_0+E_p^{\pm}} + 
\frac{\gamma_0\tilde \Lambda_{-}}{p_0-E_p^{\mp}}) \Delta \gamma_5 ]^{-1} \nonumber \\
& = & \frac{p_0-E_p^{\pm}}{p_0^2-\left({E_p^{\pm}}\right)^2-\Delta^2}\gamma_0\tilde\Lambda_{+}
+ \frac{p_0+E_p^{\mp}}{p_0^2-\left({E_p^{\mp}}\right)^2-\Delta^2}\gamma_0\tilde\Lambda_{-}\ , 
\end{eqnarray}
and
\begin{eqnarray}
\label{cosiagain}
\Xi^{\pm}=\frac{\Delta^{\pm}}{p_0^2-\left({E_p^{\pm}}\right)^2-\Delta^2}\tilde\Lambda_{+}
+\frac{\Delta^{\pm}}{p_0^2-\left({E_p^{\mp}}\right)^2-\Delta^2}\tilde\Lambda_{-}\ .
\end{eqnarray}

We can see that the Nambu-Gorkov propagator expressed in Eqs. (\ref{gpmagain}) and (\ref{cosiagain}) is very
similar to that in Eq. (\ref{limitm}). This means that we can apply the Green-function or Feynman
diagrammatic methods at moderate baryon density like that at very high baryon density.

As an application, the chiral condensate and diquark condensate can be calculated directly from the 
one-quark loop. The diquark condensate is generally expressed as
\begin{eqnarray}
< {\bar q}^C \gamma_5 q>
=-iT\sum_n \int\frac{d^3p}{(2\pi)^3}tr[\Xi^{-} \gamma_5],
\end{eqnarray}
and for the third colored quark, its chiral condensate can be evaluated by using the
quark propagator in the normal phase,
\begin{eqnarray}
<{\bar q}_3 q^3>  =   -iT \sum_n \int\frac{d^3p}{(2\pi)^3}tr[G_0^{+}] ,
\end{eqnarray}
while for the quarks participating in the diquark condensate, the chiral condensate should be
evaluated by using the quark propagator in the color superconducting phase,
\begin{eqnarray}
<{\bar q}_{1} q^{1}>  = -iT \sum_n \int\frac{d^3p}{(2\pi)^3}tr[G^{+}].
\end{eqnarray}

Using the quark propagator expressed in Eqs. (\ref{gpmagain}) and (\ref{cosiagain}),
we can also discuss the meson and diquark excitations by calculating the bubble
diagrams, and study all kinds of physical processes to investigate the
influences induced by color superconductivity.

In conclusion, we have evaluated  a more general expression for
the Nambu-Gorkov propagator 
with a mass term involved in the color-superconducting phase. 
This may help us to apply the diagrammatic method to the moderate 
density region like that in the very high density region.
The application of the propogator at moderate baryon densities  
will be described in detail in \cite{mei}, and the massive $s$ quark 
will be considered in the near future. 

Furthermore, the massive quark propagator we derived here also be useful at
high densities, where the current quark masses, especially 
for $s$ quark, could not be assumed to be zero comparing with the chemical potential.
Till now, the mass term in this region is treated perturbatively around massless limit, and the
perturbation expansion is very complicated, see \cite{rho2}. Using the massive propagator 
will simplify their expressions very much. 

\section*{Acknowledgements}
One of the authors (M.H.) thanks valuable discussions with Dr. Qishu Yan, and
thanks Prof. D. Rischke, Prof. T. Schaefer, and Dr. T. Fugleberg for reading 
the manuscript and giving valuable comments and suggestions. 
This work was supported in part by China Postdoctoral Science Foundation, 
the NSFC under Grant No. 10105005, 10135030 and 19925519,  and the 
Major State Basic Research Development Program under Contract No. G2000077407.

\end{document}